\documentclass[twocolumn,nofootinbib,amsmath,amssymb,aps,prd,balancelastpage,superscriptaddress]{revtex4-1}

\usepackage[english]{babel}
\usepackage[utf8]{inputenc}
\usepackage[colorinlistoftodos, color=green!40, prependcaption]{todonotes}
\usepackage{amsmath}
\usepackage{amsfonts}
\usepackage{amssymb}
\usepackage{dsfont}
\makeatletter\AtBeginDocument{\let\@elt\relax}\makeatother
\usepackage{graphicx}
\usepackage{bbold}

\usepackage[colorlinks=true,linkcolor=blue,urlcolor=blue,citecolor=blue]{hyperref}
\usepackage[normalem]{ulem}

\begin{document}

\title{Stochastic Ricci Flow dynamics of the gravitationally induced wave-function collapse}

\author{Matteo Lulli}
\email{lulli@sustech.edu.cn}
\affiliation{Department of Mechanics and Aerospace Engineering, Southern University of Science and Technology, Shenzhen, Guangdong 518055, China}

\author{Antonino Marcian\`o}
\email{marciano@fudan.edu.cn}
\affiliation{Center for Field Theory and Particle Physics \& Department of Physics, Fudan University, 200433 Shanghai, China}
\affiliation{Laboratori Nazionali di Frascati INFN, Frascati (Rome), Italy, EU}
\affiliation{INFN sezione Roma Tor Vergata, I-00133 Rome, Italy, EU}

\author{Kristian Piscicchia}
\email{kristian.piscicchia@cref.it}
\affiliation{Centro Ricerche Enrico Fermi - Museo Storico della Fisica e Centro Studi e Ricerche ``Enrico Fermi'', Via Panisperna, 89a, 00184, Rome, Italy, EU}
\affiliation{Laboratori Nazionali di Frascati INFN, Frascati (Rome), Italy, EU}

\begin{abstract}
\noindent
In order to reconcile the wave-function collapse in quantum mechanics with the finiteness of signals' propagation in general relativity, we delve into a stochastic version of the Ricci flow and study its non-relativistic limit in presence of matter. We hence derive the Di\'osi-Penrose collapse model for the wave-function of a quantum gas. The procedure entails additional parameters with respect to phenomenological models hitherto accounted for, including the temperature of the gas and the cosmological constant, in turn related to the stochastic gravitational noise responsible for the collapse. 
\end{abstract}

\maketitle

{\it Introduction.} --- Reconciling the local relativistic symmetries of general relativity \cite{Baez:1995sj} with the measurement problem in quantum mechanics \cite{leggett1980macroscopic,weinberg1989precision,bell2004speakable,weinberg2012collapse} is a longstanding problem, which extends to the foundational aspects of the currently accepted pillars of theoretical physics. Here we provide a natural solution to this problem, which concretely concerns the wave function collapse in quantum mechanics. The gravitational decoherence master equation, first derived by Di\'osi 
\cite{Diosi_1987,diosi1989models}, is retrieved, as the non-relativistic limit of a Quantum Gravity (QG) model, based on stochastic quantization \cite{Parisi:1980ys} and a correspondent stochastic version of the Ricci flow \cite{Hamilton_1982, Hamilton_1986, Hamilton_1993, Hamilton_1995, Perelman_entropy}. The dynamical breakdown of the superposition principle emerges, for the first time, in a fully covariant framework, hinging toward the reconciliation among quantum mechanics and general relativity. 

The mechanism that determines the departure from the quantum realm, and hence the onset of the classical behavior, has been debated since the inception of the Quantum Theory (QT) itself.  The spontaneous disappearance of quantum superposition, deeply rooted in the conundrum of measurement \cite{schrodinger1935gegenwartige}, was addressed to the friction among the basic postulates of the QT \cite{bassi2003dynamical}, specifically the linear and deterministic features of the Schr\"odinger equation, compared to the non-linear and stochastic nature of the wave-packet reduction principle. 
On that basis, an intense theoretical effort was devoted, in the last decades, to develop a quantum mechanical model of measurement capable to account for the macro-objectification, and to preserve quantum mechanical predictions for microscopic systems. Unlike hidden variable theories, which invoke the incompleteness of the Hilbert space description \cite{bohm1952suggested,bohm1989non,holland1993quantum,durr1992quantum}, the program of the Dynamical Reduction Models (DRMs) consist in introducing non-linear and stochastic modifications to the Schr\"odinger dynamics. In order to induce the localization at the wave function level and, at the same time, avoid faster than light signalling, both ingredients, namely non-linearity and stochasticity, are necessary --- see e.g. \cite{bassi2003dynamical}. Two major DRM streams dealing with this issue can be identified. On one side, Continuous Spontaneous Localization models introduce a scale that sets, through a mathematically consistent phenomenological modification of the QT, the emergence of classicality \cite{ghirardi1986unified,pearle1976reduction,pearle1979toward,pearle1989combining,ghirardi1990markov,torovs2017colored,smirne2015dissipative,adler2007collapse,adler2008collapse,di2023linear}. On the other, wave function collapse is ascribed to gravitational decoherence \cite{karolyhazy1966gravitation,karolyhazy1982possibility,karolyhazy1986possible,DIOSI1984199,Diosi_1987,diosi1989models,Penrose:1996cv,Penrose:2014nha,milburn1991intrinsic,milburn2006lorentz,adler2004quantum,adler2016gravitation,tilloy2016sourcing,bassi2017gravitational,donadi2022seven}. Despite significant recent attempts \cite{tumulka2006relativistic,tumulka2006spontaneous,bedingham2011relativistic,bedingham2014matter,pearle2015relativistic}, a fully consistent special relativistic formulation of the DRMs is still missing \cite{jones2021mass}. 

From a different viewpoint, a lot of work is being dedicated to investigate the loss of coherence of quantum matter fields, due to the interaction with a random gravitational field. Several proposals exist for quantum gravity induced  space-time fluctuations, causing decoherence. We refer to Refs. \cite{bassi2017gravitational,anastopoulos2022gravitational} for a detailed discussion. 

The role of space-time uncertainty in inducing quantum decoherence was envisaged by Feynman \cite{feynman2018feynman}, and elaborated in the last decades by many authors, the models developed by Di\'osi \cite{Diosi_1987,diosi1989models} and Penrose \cite{Penrose:1996cv,Penrose:2014nha} being the most studied ones. Several approaches share the underlying role of time stochasticity \cite{diosi2005intrinsic}, although they still differ for the grounding mechanism, the mathematical structure and the interpretation. Randomness of global (or local) time induces decoherence of the superposition of total (or local) energy eigenstates. The effect is an exponential decay of the off-diagonal terms of the density matrix, with a certain decoherence time-scale that is characteristic of the model. The corresponding structure of the master equation is peculiar, containing a double commutator term, in addition to the standard commutator in the right hand side of the von Neumann equation \cite{diosi1985orthogonal}.     \\  

The first master equation describing gravitational decoherence was developed by Di\'osi \cite{Diosi_1987}. An ultimate uncertainty is attributed to the Newtonian gravitational field \cite{Diosi_1987, Diosi:1985xp} and the correlation of local time uncertainties is derived --- in the Newtonian limit --- from the fluctuations of the gravitational potential $\phi$. The potential $\phi$ is regarded as a stochastic variable whose average coincides with the Newton potential. The master equation implies decoherence in the position basis, with the related inverse \emph{collapse time} being proportional to the gravitational self-energy of the difference between the two mass-distributions of a quantum superposition. 

Penrose's arguments are based on the uncertainty in the definition of the time-translation operator \cite{Penrose:1996cv} and on possible violations of the equivalence principle \cite{Penrose:2014nha}. These lead to the conclusion that a massive object, in a spacial superposition, gives rise to the superposition of different space-time metrics, which is unstable and has to \emph{decay}. In the non-relativistic and weak-gravitational-field limits, the expected decay time corresponds, up to a numerical factor, to the collapse time obtained by Di\'osi.   \\

Bearing these lessons in mind, stochastic quantization has been implemented in our investigation along the lines of the seminal work on gauge fields accomplished by Parisi and Wu \cite{Parisi:1980ys}. Their proposed scheme is customarily accounted as a quantization procedure alternative to canonical second quantization and path integral quantization, which takes inspiration from relaxation processes in complex systems. Specifically, the stochastic quantization method is implemented by a Langevin equation, the drift term of which is represented by the Euler-Lagrange equations of motion, and with stochastic noise relative to the field to be assigned. This realizes a heat-kernel like evolution in the thermal time that drives the system toward equilibrium, where the equations of motion hold and the standard symmetries are recovered. In the out-of-equilibrium phase, divergencies are associated to symmetry-violating terms within the fields propagators~\cite{Parisi:1980ys}. \\ 

{\it Preliminaries.} --- The stochastic completion of the Ricci flow implements the Langevin equation for the gravitational field \cite{Lulli:2021bme}. This realizes a conceptual subtlety: being the symmetries of the system broken in the out-of-equilibrium phase, stochastic quantization prevents from implementing as strong operatorial equations the standard symmetry-constraints of gravity, which are rather recovered only at equilibrium, as classical equations of motion. Therefore, the Stochastic Ricci Flow (SRF) provides a dynamical description (in the thermal time) of the breakdown of the symmetries, providing a paradigm shift in the quantization methods. Furthermore, SRF provides the RG flow of the gravitational field, setting the hierarchy of scales, dictating their dynamical evolution in the thermal time\footnote{The thermal time $s$, often improperly called ``stochastic time'', can be represented as an affine parameter whose integral transforms as a scalar under diffeomorphisms. This is consistent with results of Stochastic Quantization of two-dimensional causal dynamic triangulations --- see e.g. \cite{Ambjorn:2009wi} --- in which the proper time $\tau$ acts as a thermal time. In higher dimensional theories the proper-time dependence is encoded in the coordinate dependence, with the total proper-time derivative being expressed by $\ell {\mbox{d}}/{\mbox{d}\tau} = u^\mu\, \nabla_\mu$, where $u^\mu$ denotes a time-like vector and $\ell$ a length constant. The identification $u^\mu = n^\mu$ then introduces a space-like foliation. The thermal time $s$ and the proper time $\tau$ are connected through the  identification of $\mbox{d}x^\mu/\mbox{d}s = n^\mu$, the covariant derivative of the scalar time field $n_\mu = -N\nabla_\mu t$, from which it follows that $\varepsilon(s) \delta s = -N \delta t$, with $\varepsilon = n^\mu n_\mu$.}. The equilibrium limit corresponds to the IR fixed point of SRF. Nonetheless, the gravitational theory might flow toward a UV fixed point, a feature linking SRF to Ho\u rava-Lifshitz gravity \cite{Frenkel:2020djn,Frenkel:2020dic,Frenkel:2020ixs}. It has been proved in \cite{Frenkel:2020djn}, in a super-symmetric setting implied by the BRST symmetry, that the RG flow of Ho\u rava-Lifshitz gravity is indeed the Ricci flow, with a topological theory in the UV fixed point.

Rumpf was the first author to study the implementation of stochastic quantization methods to Einstein gravity, introducing in \cite{Rumpf:1985eh} a one-parameter family of covariant Langevin equations for the metric $g_{\mu \nu}(s;x^\alpha)$, namely a DeWitt one-parameter family of super-metrics. Following Rumpf's notation, the Langevin equation reads 
\begin{equation}
\frac{\partial g_{\mu \nu}}{\partial s}= - 2 \imath \left( R_{\mu \nu} - \frac{\lambda +1}{2(2\lambda +1)} g_{\mu \nu} R\right) +\eta_{\mu \nu}
\,,
\end{equation}
with $\lambda$ a parameter labelling the supermetric $\mathcal{G}_{\mu \nu \rho\sigma}$ that interpolates, within the drift term, among the Ricci ($\lambda=-1$) and the Einstein ($\lambda=0$) tensors, and $\eta_{\mu \nu}$ the stochastic noise for the 
metric field $g_{\mu \nu}(s; x^\alpha)$. It has been argued in Ref.~\cite{Rumpf:1985eh} that the choice $\lambda=-1$, which implies the Ricci flow, corresponds to the DeWitt's choice of the measure (in the path-integral approach to the quantization of gravity) as invariant under general coordinate transformations. Ref.~\cite{Rumpf:1985eh} has been also crucial in stating the choice of the It\^o calculus with respect to the Stratonovich one, which is 
related to form-invariance of the viel-bein functional in the super-field space.     

Essential for the development of our mechanism is the choice of a multiplicative stochastic gravitational noise, namely $\eta_{\mu \nu}=\eta g_{\mu \nu}$. 
%
%
From a physical perspective, the scalar nature of the noise, with the dimensions $[L^{-2}]$, provides an intuitive connection with the idea of thermal curvature fluctuations. Furthermore, one can draw a parallel with stochastic hydrodynamics~\cite{LANDAU_1992}: the non-linear formulation~\cite{Kim_1991} naturally yields multiplicative noise. Hence, we argue that the choice of a multiplicative noise should be appropriate for the non-linear gravitational regime while recovering the additive case in the linearisation against a Minkowski background.
Within this choice, and expressing the drift term in its dependence from the variational principle and the supermetric $\mathcal{G}_{\mu \nu \rho\sigma}$, the Langevin equation for the metric field $g_{\mu \nu}$ can be recast as 
\begin{equation}
 \frac{\partial g_{\mu \nu}}{\partial s}= \imath \, \mathcal{G}_{\alpha \beta \mu \nu} \, \frac{\delta S}{\delta g_{\alpha \beta}} + g_{\mu \nu} \eta \,.
\end{equation}  
The related Fokker-Planck equation for the probability distribution $p$ can be consistently recovered accounting for the It\^o calculus:
\begin{equation}
\frac{\partial p}{\partial s}= -\left[ \mathcal{G}_{\alpha \beta \mu \nu} \, \frac{\delta S}{\delta g_{\alpha \beta}} \right] + \frac{\delta^2}{\delta g_{\mu \nu} \delta g_{\alpha \beta}} \left[ g_{\mu \nu} g_{\alpha \beta} \, p \right]\,.
\end{equation} 
It was already observed in \cite{Lulli:2021bme} that an approximate stationary  solution of the Fokker-Planck equation is provided by the expression
\begin{equation}
p\simeq \frac{D}{g^2_{\mu \nu}} \exp \left[ 2 \int^{g_{\mu \nu}} \mathcal{D}g_{\alpha \beta} \, \frac{
\mathcal{G}_{\alpha \beta \mu \nu}  \, \frac{\delta S}{\delta g_{\alpha \beta}}
}{\Lambda_0 g^2_{\alpha \beta}}
\right]
\,, 
\end{equation} 
which is extremized by the equations 
\begin{equation}
\mathcal{G}_{\alpha \beta \mu \nu} \, \frac{\delta S}{\delta g_{\alpha \beta}} - \imath \Lambda_0 g_{\mu \nu}=0\,.
\end{equation}
For a complex noise $\eta$ specified by the relation $\eta=\sigma_{\tilde{\eta}} \tilde{\eta}$, with $\tilde{\eta}$ real and $\sigma_{\tilde{\eta}}=\sqrt{\Lambda}_0$, a rotation of the stochastic noise on the complex plane by $\pi/4$, namely $\sigma_{\tilde{\eta}} \rightarrow e^{-\imath \frac{\pi}{4}} \sigma_{\tilde{\eta}} $, provides the interpretation of $\sigma^2_{\tilde{\eta}}$, the square of the 
amplitude of the stochastic noise, as cosmological constant. This identifies the origin of the stochastic fluctuations of space-time with the cosmological constant.  \\  


{\it Derivation of the wave-function collapse.} --- In order to derive the collapse of the wave-function of a quantum system we consider the non-relativistic limit of the metric tensor, in its diagonal form and dependence only from the lapse function \cite{Arnowitt:1959ah}. Thus the only non-flat term of the metric tensor is expressed by $N=\sqrt{-g_{00}}$, which satisfies 
\begin{equation}
\frac{\partial N}{\partial g_{00}}=-\frac{1}{2\sqrt{-g_{00}}}=-\frac{1}{2N}\,. \label{g00}
\end{equation}
Following the rules of the It\^o calculus for a generic variable transformation $N=N(g_{\mu\nu})$ and setting the noise amplitude in the generic form $\sigma_{\tilde{\eta}}=e^{\imath\gamma/2}\sqrt{2\alpha}$, we can express the Langevin equation in terms of the new variables, i.e.
\begin{eqnarray}
    \frac{\partial N}{\partial s}= 
    &&-2\imath\frac{\partial N}{\partial g_{\mu\nu}}\left[R_{\mu\nu}-R_{\mu\nu}^{\text{T}}\right]+\alpha e^{\imath\gamma}g_{\mu\nu}g_{\alpha\beta}\frac{\partial^{2}N}{\partial g_{\mu\nu}\partial g_{\alpha\beta}} \nonumber \\ 
    && + e^{\imath\gamma/2}\sqrt{2\alpha}\frac{\partial N}{\partial g_{\mu\nu}}g_{\mu\nu}\tilde{\eta}\,,
\end{eqnarray}
in which the target Ricci tensor $R_{\mu\nu}^{\text{T}}$ has been introduced, defined in terms of the energy-momentum tensor $T_{\mu \nu}$
\begin{equation}
R_{\mu\nu}^{\text{T}}=T_{\mu \nu} - \frac{T}{2}  g_{\mu \nu}\,,
\end{equation}
to recover in the vanishing noise limit and asymptotically, at stationarity in the thermal time $s$, the Einstein equations (without cosmological constant).

Implementing now the transformation in Eq.~\eqref{g00}, the stochastic differential equation (SDE) rewrites 
\begin{equation}
\frac{\partial N}{\partial s}  =\frac{\imath}{N}\left[R_{00}-R_{00}^{\text{T}}\right]-\frac{\alpha}{4}e^{\imath\gamma}N+e^{\imath\gamma/2}\sqrt{\frac{\alpha}{2}}N\tilde{\eta}\,.
\end{equation}

We introduce matter as an ideal (non-interacting) gas, with isotropic pressure expressed in terms of the temperature $T$ as
\begin{equation} \label{piro}
    p=\rho c_{s}^{2}=\rho\frac{\gamma k_{B}T}{m}=\rho\sigma T,\qquad {\rm with} \qquad \sigma=\frac{\gamma k_{B}}{m},
\end{equation}
$c_s^2$ denoting the square of the speed of sound, $\gamma$ is the polytropic factor, $m$ is the molecular mass and $k_B$ is the Boltzmann constant. The stress-energy tensor finally reads 
\begin{equation}
 T_{\,\nu}^{\mu}=\text{diag}\left( -\rho ,\rho\sigma T,\rho\sigma T,\rho\sigma T\right)\,.
\end{equation}
The covariant time-time component and the trace can be written as
\begin{equation}
T_{00}=g_{0\mu}T_{\,0}^{\mu}\!=\!N^{2}\rho \,, \ \ T=T_{\:\nu}^{\mu}\delta_{\,\mu}^{\nu}
    \!=\! \rho\left(3\sigma T-1\right)\!. \ \ \
\end{equation}
This implies the expression of the target Ricci tensor 
\begin{equation}
 R_{00}^{\text{T}}=\kappa\left(T_{00}-\frac{1}{2}g_{00}T\right) = \frac{\kappa}{2}N^{2}\rho\left(3\sigma T+1\right)\,.
\end{equation}
Assuming space-time to be static, the time-time component of the Ricci tensor can be expressed as
\begin{equation}
    R_{00}\simeq -\, \partial_{\alpha}\Gamma_{00}^{\alpha}=\frac{1}{2}\partial_{\alpha}\partial^{\alpha}g_{00}
    =-\frac{1}{2}\nabla^{2}N^{2}\,.
\end{equation}
Inserting the definitions from Eq.~\eqref{piro}, we can recast the SDE for the lapse function as
\begin{eqnarray}
    \frac{\partial N}{\partial s}=&-& \frac{\imath}{2N}\left[\nabla^{2}N^{2}+\kappa N^{2}\rho\left(3\sigma T+1\right)\right] \nonumber \\
    &-& \frac{\alpha}{4}e^{\imath\gamma}N+e^{\imath\gamma/2}\sqrt{\frac{\alpha}{2}}N\tilde{\eta} \,.
\end{eqnarray}
The associated Fokker-Planck equation for the probability distribution $p[N]$ can be recovered --- see e.g. \cite{gardiner2004handbook} --- first expressing the SDE in the differential form
\begin{eqnarray} \label{dN}
    \text{d}N= &-& \left\{ \frac{\imath}{2N}\left[\nabla^{2}N^{2}+\kappa N^{2}\rho\left(3\sigma T+1\right)\right]+\frac{\alpha}{4}e^{\imath\gamma}N\right\} \text{d}s \nonumber\\
 &+&e^{\imath\gamma/2}\sqrt{\frac{\alpha}{2}}N\text{d}W \equiv a \text{d}s  + b  \text{d}W \,,
\end{eqnarray}
where $\mbox{d}W$ is the Wiener process, and then extending it to a generic functional $f$ of $N$ --- in the last equality of Eq.~\eqref{dN} we have introduced the definition of $a$ and $b$. We may now derive up to the second order the stochastic differential of $f$, and then evaluate the expected value of the derivative $\langle\mbox{d}f/\mbox{d}s\rangle$, using its explicit definition and the time derivative of the probability distribution. Equating the two results, we finally recover the Fokker-Planck equation.

Functional derivatives are performed within these passages, introducing an integration over test functions, while the singular nature of two consecutive functional derivations is taken into account by introducing a non-trivial kernel $k_{\varepsilon}\left(x,y\right)$ for the noise correlation function, which converges to a Dirac delta in some limit $\varepsilon\to0$. 

After several manipulations --- see e.g. appendix \ref{FPE} ---  and using the property $k_{\varepsilon}\left(x,x\right)=2\alpha^2$, we derive the Fokker-Planck equation 
\begin{equation}
a_{x}p\left[N_{x}\right]-\frac{e^{\imath\gamma}\alpha}{4}\left(2n_{x}\,N_{x}p\left[N_{x}\right]+n_{x}N_{x}^{2}\frac{\delta}{\delta N_{x}}p\left[N_{x}\right]\right)=0\,,
\end{equation}
in which $ n(x)\equiv n_x $ is a test function. An approximated solution to the Fokker-Planck equation can be recovered --- see appendix \ref{FPE} for details --- that is expressed as 
\begin{eqnarray}
p\left[N_{x}\right] &=& B\exp\left\{ \int^{N_{x}}\prod_{y}\text{d}N_{y}\frac{4}{e^{\imath\gamma}\alpha N_{y}^{2}}\left[a_{y}-\frac{e^{\imath\gamma}\alpha}{2}\,N_{y}\right]\right\}  \nonumber\\ 
&=&B\exp\left[\frac{\imath}{\hbar}S_{0}\right]\,, \label{pNp}
\end{eqnarray}
with $B$ integration constant, and is such that in the {\it non-relativistic limit} it corresponds to the exponential of the action $S_0$, according to a general result that holds for the Euclidean path-integral measure.

Not taking into account the steady-state limit --- the semi-classical solution will be substituted at the end of the calculation, hence casting the action at the equilibrium limit --- we may proceed to write a wave-functional $\Psi[N]$ in the WKB approximation, finding 
\begin{equation}
 \Psi=\exp\left[\frac{\imath}{\hbar}S_{0}\right]\,,
\end{equation}
with $S_0$ specified in Eq.~\eqref{pNp}.
The stochastic equation for the wave-functional is recovered to be
\begin{eqnarray}
\!\!\!\!\!\!\!\!\!\!\!\! \text{d}\Psi \! &=& \!\! \int\!\! \text{d}^{4}x\,m_{x}\,\frac{\delta\Psi}{\delta N_{x}}a_{x}\text{d}s+\int\text{d}^{4}x\,m_{x}\,\frac{\delta\Psi}{\delta N_{x}}b_{x}\text{d}W_{x} \nonumber \\
\!\!\!& +& \!\frac{1}{2}\!\int\!\text{d}^{4}x \, m_{x}\!\! \! \int\!\!\text{d}^{4}y\, n_{y}\,b_{x}b_{y}\,k_{\varepsilon}\left(x,y\right)\frac{\delta}{\delta N_{x}}\frac{\delta}{\delta N_{y}}\Psi\text{d}s.
\end{eqnarray}


{\it Non-relativistic limit and Di\'osi-Penrose models.} --- The differential of the wave-functional can be finally determined --- see appendices \ref{FPE} and \ref{NRL}. The solution, within the non-relativistic limit  
\begin{equation}
    g_{00}=-1-2\phi,\qquad N=\sqrt{1+2 \phi}\simeq1+\phi\,,
\end{equation}
and within the approximations $\nabla^2 N\simeq \nabla^2 \phi=\kappa \rho /2$, which makes use of the classical limit, and 
$N\simeq 1$, provides the final expression
\begin{eqnarray} \label{collapse}
\text{d} \Psi \!&=&\! \left\{\int\text{d}^{4}x\,m_{x}\,\left[2\imath\kappa\rho_{x}\left(2+3\sigma T\right)+\frac{\alpha}{4}e^{\imath\gamma}\right]\text{d}s \right\}\, \Psi \\
&-& \left\{    \frac{e^{-\imath\frac{\gamma}{2}}}{\sqrt{2 \alpha}} \!\! \int \!\! \text{d}^{4}x\,m_{x}\! \left[2\imath\kappa \rho_{x}\left(3\sigma T+2\right) \!+\! 3\alpha e^{\imath\gamma}\right] \! \text{d}W_{x} \! \right\} \! \Psi \nonumber\\
&+& \Big\{\frac{1}{2}\int\text{d}^{4}x\,m_{x}\,\int\text{d}^{4}y\,n_{y}\,\frac{1}{2}\,k_{\varepsilon}\left(x,y\right)\frac{1}{e^{\imath\gamma}\alpha}\text{d}s \times \nonumber \\
 & \phantom{a}& \!\!\!\!\!\!\! \left[2 \imath \kappa\rho_{x}\left(2\!+\! 3\sigma T\right)\!+\! 3 \alpha e^{\imath\gamma}\right] 
  \left[2 \imath \kappa\rho_{y}\left(2\!+\! 3\sigma T\right)\!+\!3 \alpha e^{\imath\gamma}\right] 
 \! \Big\}\Psi. \nonumber
\end{eqnarray}

{\it Gorini-Kossakowski-Lindblad-Sudarshan equation.} --- The components of Eq.~\eqref{collapse} have been studied in the semi-classical limit, for which quantities are intended to be commutative. We may take into account now two known facts: i) the thermal time is monotonically dependent on the proper time, as shown in \cite{Lulli:2021bme} and reminded previously, thus a transformation among the two can be recovered
; ii) Eq.~\eqref{collapse} can be represented, after first quantization, on the phase space of a quantum system under scrutiny, and hence regarded as a deformation of the Hamiltonian time-flow in quantum mechanics. A correspondent operatorial equation can be derived, which connects to the Gorini-Kossakowski-Lindblad-Sudarshan (GKLS) equation \cite{GKS,Lind}, retaining the general form of a Markovian master equation that describes open quantum systems, along the prescriptions specified in \cite{DaviesA, DaviesB}, namely 
\begin{eqnarray} \label{GenGKLS}
\frac{\text{d}}{\text{d} s}\rho_\Psi &\equiv&  \left\langle \frac{\text{d}}{\text{d} s}\rho_\Psi \right\rangle_{\tilde{\eta}} =  \imath \int d^4x \, m_x \left[ \mathcal{H}_x, \rho_\Psi \right]  \\
&&+ \int d^4x\,  m_x  \int d^4y\,  n_y \, \widetilde{k}_{\varepsilon}\left(x,y\right) \, \mathcal{L}_x \rho_\Psi \mathcal{L}_y^\dagger \nonumber\\
&&-\frac{1}{2}  \int d^4x\,  m_x  \int d^4y\,  n_y \, \widetilde{k}_{\varepsilon}\left(x,y\right)\,  \{\mathcal{L}_x^\dagger \mathcal{L}_y, \rho_\Psi \}\,,  \nonumber
\end{eqnarray} 
where the density matrix related to a quantum state $|\Psi\rangle$, in turn related to the wave-functional $\Psi$, is denoted as $\rho_\Psi$ ---  see appendix \ref{SecGKLS} for the details of the derivation. Eq.~\eqref{GenGKLS} entails a generalized Schr\"odinger equation for open quantum systems, in which decohering effects are provided by a thermal bath --- in our case, the stochastic noise forming a cosmological constant background. Eq.~\eqref{GenGKLS} generates a dynamics that, although being not unitary, is still trace-preserving and positive for any initial condition. It represents a generalization of the GKLS form for a non-gaussian stochastic noise characterized by a specific choice of the kernel $k_{\varepsilon}\left(x,y\right)$ --- see appendix \ref{SecGKLS}. \\

{\it Discussion.} --- The low temperature limit can be inspected dropping the $\sigma T$ terms out from Eq.~\eqref{collapse}. The Di\'osi-Penrose collapse models are achieved through: i) a colored-noise 
choice of the kernel $k_{\varepsilon}\left(x,y\right)$ that corresponds to a white noise over the time axis multiplied by the Newtonian potential over space; ii) the inclusion in the kernel $k_{\varepsilon}\left(x,y\right)$, for dimensional reasons, of the parameter $\alpha$, connected to the square of the amplitude of the gravitational stochastic noise and proportional to the cosmological constant. The stochastic noise of the gravitational field is at the origin of the decohering effects responsible for the gravitationally-induced collapse of the wave-function. Stochasticity of the noise, in the SRF approach, can also be intended as the inherent fuzziness from which space-time emerges \cite{Liu:2023pok}, as well as the source of scrambling of the information \cite{Su:2023ciz,Su:2023teb}.   \\

{\it Conclusions.} --- The non-relativistic limit derived from the SRF approach on one side hinges toward a reconciliation of the local relativistic symmetries of general relativity with the measurement problem in quantum mechanics; on the other, it introduces a wider class of collapse models than the Di\'osi-Penrose class, as it accounts for both the temperature and the cosmological-constant dependence of decoherence effects. This ultimately suggests a possible multi-messenger approach toward the phenomenology of the collapse of the gravitational wave-function of quantum systems, in which the class of models can be falsified also accounting for astrophysical constraints. 

The novel SRF approach also enables to recover a solid theoretical link between models of quantum gravity (and related universality classes) and their consequences for the foundational aspects of quantum mechanics that can be experimentally falsified \cite{pro}, hence prospecting a new way to develop quantum gravity phenomenology. This represents a promising perspective for forthcoming developments in the field of quantum gravity phenomenology.   

%
\section*{Acknowledgments}
\noindent 
We thank Catalina Curceanu, Lajos Di\'osi, Chris Field, Felix Finster, James Glazebrook, Ugo Moschella and Claudio Paganini for useful comments and suggestions.
A.M.\ acknowledges support by the NSFC, through the grant No.\ 11875113, the Shanghai Municipality, through the grant No.\ KBH1512299, and by Fudan University, through the grant No.\ JJH1512105.
K.P.\ acknowledges the support of Grant 62099 from the John Templeton Foundation. The opinions expressed in this publication are those of the authors and do not necessarily reflect the views of the John Templeton Foundation.
K.P.\ acknowledges support from the Centro Ricerche Enrico Fermi - Museo Storico della Fisica e Centro Studi e Ricerche ``Enrico Fermi'' (Open Problems in Quantum Mechanics project).

\onecolumngrid
\appendix
\section{It\^o Calculus} \label{Aa}
\noindent
Throughout this work, we adopt the signs' convention proposed by Misner, Thorne and Wheeler \cite{MTW_1973}, namely $([S1],[S2],[S3])=(+,+,+)$. We denote $\kappa = 8\pi G / c^4$.

\subsection{Conventions and notations}
\noindent 
Following the rules of It\^o calculus \cite{gardiner2004handbook} for a generic transformation of variables $N=N(g_{\mu\nu})$, we can write the new stochastic differential equation (SDE) as
\begin{equation}
    \frac{\partial N}{\partial s}=-2\imath\frac{\partial N}{\partial g_{\mu\nu}}\left[R_{\mu\nu}-R_{\mu\nu}^{\text{T}}\right]+\alpha e^{\imath\gamma}g_{\mu\nu}g_{\alpha\beta}\frac{\partial^{2}N}{\partial g_{\mu\nu}\partial g_{\alpha\beta}}+e^{\imath\gamma/2}\sqrt{2\alpha}\frac{\partial N}{\partial g_{\mu\nu}}g_{\mu\nu}\tilde{\eta}\,,
\end{equation}
having introduced, as in \cite{Lulli:2021bme}, the definition 
\begin{equation}
\eta=\sigma_{\tilde{\eta}}\, \tilde{\eta}  = \sqrt{2\alpha}\, e^{\imath\gamma/2} \, \tilde{\eta} \,,
\end{equation}
with $\alpha$ real and positive constant, with physical dimension of energy square, identified with the cosmological constant, and $\gamma$ a phase.\\ 

Choosing now $N$ to be the lapse function of a diagonal metric tensor, i.e. $N=\sqrt{-g_{00}}$, the first and the second derivative can be reshuffled into
\begin{equation}
    \frac{\partial N}{\partial g_{00}}=-\frac{1}{2\sqrt{-g_{00}}}=-\frac{1}{2N}   
\end{equation}
and
\begin{equation}
    \frac{\partial^{2}N}{\partial g_{00}^{2}}=-\frac{1}{2}\frac{\partial}{\partial g_{00}}\left[\frac{1}{\sqrt{-g_{00}}}\right]=-\frac{1}{2}\left[-\frac{1}{2}\frac{1}{\left(-g_{00}\right)^{3/2}}\frac{\partial\left(-g_{00}\right)}{\partial g_{00}}\right]=-\frac{1}{4}\frac{1}{\left(-g_{00}\right)^{3/2}}=-\frac{1}{4N^{3}}\,.
\end{equation}
Consequently, we may find that
\begin{equation}
\begin{split}
\frac{\partial N}{\partial s} & =-2\imath\frac{\partial N}{\partial g_{00}}\left[R_{00}-R_{00}^{\text{T}}\right]+\alpha e^{\imath\gamma}\left(g_{00}\right)^{2}\frac{\partial^{2}N}{\partial g_{00}^{2}}+e^{\imath\gamma/2}\sqrt{2\alpha}\frac{\partial N}{\partial g_{00}}g_{00}\tilde{\eta}\\
 & =\frac{\imath}{N}\left[R_{00}-R_{00}^{\text{T}}\right]-\frac{\alpha e^{\imath\gamma}N^{4}}{4N^{3}}+e^{\imath\gamma/2}\sqrt{2\alpha}\left(-\frac{1}{2N}\right)\left(-N^{2}\right)\tilde{\eta}\,,\\
\end{split}
\end{equation}
and thus
\begin{equation}
\begin{split} \label{Nesse}
\frac{\partial N}{\partial s} & =\frac{\imath}{N}\left[R_{00}-R_{00}^{\text{T}}\right]-\frac{\alpha}{4}e^{\imath\gamma}N+e^{\imath\gamma/2}\sqrt{\frac{\alpha}{2}}N\tilde{\eta}\,.\\
\end{split}
\end{equation}

\section{Matter description and related energy-momentum tensor}
\noindent 
We introduce matter in the form of an ideal (non-interacting) gas, the isotropic pressure of which can be expressed in terms of the temperature $T$ as
\begin{equation}
    p=\rho c_{s}^{2}=\rho\frac{\gamma k_{B}T}{m}=\rho\sigma T,\qquad\sigma=\frac{\gamma k_{B}}{m}\,,
\end{equation}
where $c_s^2$ is the square of the speed of sound, $\gamma$ is the polytropic factor, $m$ is the molecular mass and $k_B$ is the Boltzmann constant. The stress-energy tensor reads 
\begin{equation}
T_{\,\nu}^{\mu}=\text{diag}\left( -\rho 
    ,\rho\sigma T,\rho\sigma T,\rho\sigma T\right)\,.
\end{equation}
The covariant time-time component and the trace can be written as
\begin{equation}
T_{00}=g_{0\mu}T_{\,0}^{\mu}=N^{2}\rho ,\qquad T=T_{\:\nu}^{\mu}\delta_{\,\mu}^{\nu}=-\rho +3\rho\sigma T=\rho\left(3\sigma T-1\right)\,.
\end{equation}
Consequently, the target Ricci tensor reads
\begin{equation}
R_{00}^{\text{T}}=\kappa\left(T_{00}-\frac{1}{2}g_{00}T\right)=\kappa\left(N^{2}\rho +\frac{1}{2}N^{2}\rho\left(3\sigma T-1\right)\right)=\frac{\kappa}{2}N^{2}\rho\left(3\sigma T+1\right)\,.
\end{equation}
Assuming space-time to be static, we may recast the time-time component of the Ricci tensor as
\begin{equation}
    R_{00}\simeq- \, \partial_{\alpha}\Gamma_{00}^{\alpha}=\frac{1}{2}\partial_{\alpha}\partial^{\alpha}g_{00}=\frac{1}{2}\partial_{\alpha}\partial^{\alpha}\left(-N^{2}\right)=-\frac{1}{2}\nabla^{2}N^{2}.
\end{equation}
These latter definitions can be inserted in Eq.~\eqref{Nesse}, which finally enables to derive the SDE for the lapse function as
\begin{equation} \label{Nesse2}
    \frac{\partial N}{\partial s}=-\frac{\imath}{2N}\left[\nabla^{2}N^{2}+\kappa N^{2}\rho\left(3\sigma T+1\right)\right]-\frac{\alpha}{4}e^{\imath\gamma}N+e^{\imath\gamma/2}\sqrt{\frac{\alpha}{2}}N\tilde{\eta} \,.
\end{equation}

\section{Fokker-Planck equation and WKB approximation}\label{FPE}
\noindent 
We may now re-express the SDE \eqref{Nesse2} in the differential form
\begin{equation}\label{Nesse3}
    \text{d}N=\left\{ -\frac{\imath}{2N}\left[\nabla^{2}N^{2}+\kappa N^{2}\rho\left(3\sigma T+1\right)\right]-\frac{\alpha}{4}e^{\imath\gamma}N\right\} \text{d}s+e^{\imath\gamma/2}\sqrt{\frac{\alpha}{2}}N\text{d}W\,,
\end{equation}
where $\text{d}W$ is the Wiener process. Given a generic functional $f$ of $N$, we proceed to derive the Fokker-Planck equation for the probability distribution $p[N]$ by generalizing the procedure presented in~\cite{gardiner2004handbook}. First, we derive up to the second order (in the differential of the thermal time $s$) the stochastic differential of $f$; then, we evaluate the expectation value of the derivative $\langle\mbox{d}f/\mbox{d}s\rangle$ by considering both its explicit definition and the result obtained from the time derivative of the probability distribution; finally, by equating the two results, we obtain the Fokker-Planck equation.\\

In order to compute the functional expression for $\text{d}f$, we need to take into account that the ordinary derivatives need to be substituted with the functional ones by introducing the integration over the test functions. Furthermore, we need to tame the singular nature of two consecutive functional derivations, by introducing a non-trivial kernel for the noise correlation function, converging to a Dirac delta in some limit of the parameter $\varepsilon$, i.e. $\varepsilon\to0$.\\

Considering first the stochastic differential of $N$ 
\begin{equation}
    \begin{split}
\text{d}N & =a\text{d}s+b\text{d}W\\
\\
a &= -\frac{\imath}{2N}\left[\nabla^{2}N^{2}+\kappa N^{2}\rho\left(3\sigma T+1\right)\right]-\frac{\alpha}{4}e^{\imath\gamma}N\\
b &= e^{\imath\gamma/2}\sqrt{\frac{\alpha}{2}}N\,,\\
\end{split}
\end{equation}
we may then apply the aforementioned strategy in order to recover the stochastic differential of $f[N]$, namely 
\begin{equation}
    \begin{split}
\text{d}f & \equiv f\left[N+\text{d}N\right]-f\left[N\right]\\
 & \simeq\int\text{d}^{4}x\,m_{x}\,\frac{\delta f}{\delta N_{x}}\text{d}N_{x}+\frac{1}{2}\int\text{d}^{4}x\text{d}^{4}y\,m_{x}\,n_{y}\,\text{d}N_{x}\text{d}N_{y}\frac{\delta}{\delta N_{x}}\frac{\delta}{\delta N_{y}}f\\
 & =\int\text{d}^{4}x\,m_{x}\,\frac{\delta f}{\delta N_{x}}\left(a_{x}\text{d}s+b_{x}\text{d}W_{x}\right)+\frac{1}{2}\int\text{d}^{4}x\text{d}^{4}y\,m_{x}\,n_{y}\,b_{x}b_{y}\text{d}W_{x}\text{d}W_{y}\frac{\delta}{\delta N_{x}}\frac{\delta}{\delta N_{y}}f\\
 & =\int\text{d}^{4}x\,m_{x}\,\frac{\delta f}{\delta N_{x}}\left(a_{x}\text{d}s+b_{x}\text{d}W_{x}\right)+\frac{1}{2}\int\text{d}^{4}x\text{d}^{4}y\,m_{x}\,n_{y}\,b_{x}b_{y}\left(\text{d}W_{x}\right)^{2}k_{\varepsilon}\left(x,y\right)\frac{\delta}{\delta N_{x}}\frac{\delta}{\delta N_{y}}f\\
 & =\int\text{d}^{4}x\,m_{x}\,\frac{\delta f}{\delta N_{x}}\left(a_{x}\text{d}s+b_{x}\text{d}W_{x}\right)+\frac{1}{2}\int\text{d}^{4}x\text{d}^{4}y\,m_{x}\,n_{y}\,b_{x}b_{y}\text{d}s\,k_{\varepsilon}\left(x,y\right)\frac{\delta}{\delta N_{x}}\frac{\delta}{\delta N_{y}}f\,,\\
\end{split}
\end{equation} 
in which we have introduced the test functions $n(x)$ and $m(x)$, denoted at the point $x$ respectively as $n_x$ and $m_x$. This equation finally provides the expression
 \begin{equation}
\begin{split}
\text{d}f & =\int\text{d}^{4}x\,m_{x}\,\left[\frac{\delta f}{\delta N_{x}}a_{x}+\frac{1}{2}\int\text{d}^{4}y\,n_{y}\,b_{x}b_{y}\,k_{\varepsilon}\left(x,y\right)\frac{\delta}{\delta N_{x}}\frac{\delta}{\delta N_{y}}f\right]\text{d}s+\int\text{d}^{4}x\,m_{x}\,\frac{\delta f}{\delta N_{x}}b_{x}\text{d}W_{x}\,.
\end{split}
\end{equation}
We may at this point determine the associated Fokker-Planck, starting from
\begin{equation}
\begin{split}
&\text{d}f=\int\text{d}^{4}x\,m_{x}\,\left[\frac{\delta f}{\delta N_{x}}a_{x}+\frac{1}{2}\int\text{d}^{4}y\,n_{y}\,b_{x}b_{y}\,k_{\varepsilon}\left(x,y\right)\frac{\delta}{\delta N_{x}}\frac{\delta}{\delta N_{y}}f\right]\text{d}s+\int\text{d}^{4}x\,m_{x}\,\frac{\delta f}{\delta N_{x}}b_{x}\text{d}W_{x}\,,\\
\end{split}
\end{equation}
and hence averaging over the stochastic noise probability distribution --- we denote this operation with $\left\langle \ \ \cdot \ \ \right\rangle_{\tilde{\eta}} $ --- as it follows: 
\begin{equation}
\begin{split}
&\left\langle\frac{\text{d}f}{\text{d}s}\right\rangle_{\tilde{\eta}}=\left\langle\int\text{d}^{4}x\,m_{x}\,\left[\frac{\delta f}{\delta N_{x}}a_{x}+\frac{1}{2}\int\text{d}^{4}y\,n_{y}\,b_{x}b_{y}\,k_{\varepsilon}\left(x,y\right)\frac{\delta}{\delta N_{x}}\frac{\delta}{\delta N_{y}}f\right]+\int\text{d}^{4}x\,m_{x}\,\frac{\delta f}{\delta N_{x}}b_{x}\tilde{\eta}_{x}\right\rangle_{\tilde{\eta}}\\
&=\left\langle\int\text{d}^{4}x\,m_{x}\,\left[\frac{\delta f}{\delta N_{x}}a_{x}+\frac{1}{2}\int\text{d}^{4}y\,n_{y}\,b_{x}b_{y}\,k_{\varepsilon}\left(x,y\right)\frac{\delta}{\delta N_{x}}\frac{\delta}{\delta N_{y}}f\right]\right\rangle_{\tilde{\eta}}\\
&=\int\prod_{z}\text{d}N_{z}\,p\left[N_{z},s\right]\int\text{d}^{4}x\,m_{x}\,\left[\frac{\delta f}{\delta N_{x}}a_{x}+\frac{1}{2}\int\text{d}^{4}y\,n_{y}\,b_{x}b_{y}\,k_{\varepsilon}\left(x,y\right)\frac{\delta}{\delta N_{x}}\frac{\delta}{\delta N_{y}}f\right]\\&=\int\text{d}^{4}x\,\int\prod_{z}\text{d}N_{z}\,p\left[N_{z},s\right]m_{x}\,\left[\frac{\delta f}{\delta N_{x}}a_{x}+\frac{1}{2}\int\text{d}^{4}y\,n_{y}\,b_{x}b_{y}\,k_{\varepsilon}\left(x,y\right)\frac{\delta}{\delta N_{x}}\frac{\delta}{\delta N_{y}}f\right]\,.
\end{split}
\end{equation}
Now, we perform functional integration by parts and obtain
\begin{equation}
    \begin{split}  &\left\langle\frac{\text{d}f}{\text{d}s}\right\rangle_{\tilde{\eta}}=\int\text{d}^{4}x\,\text{d}N_{x}\,m_{x}\,\left[\frac{\delta f}{\delta N_{x}}a_{x}p\left[N_{x},s\right]+\frac{1}{2}\int\text{d}^{4}y\,n_{y}\,b_{x}b_{y}p\left[N_{x},s\right]\,k_{\varepsilon}\left(x,y\right)\frac{\delta}{\delta N_{x}}\frac{\delta}{\delta N_{y}}f\right]\\&=\int\text{d}^{4}x\,\text{d}N_{x}\,m_{x}\,f\left[-\frac{\delta}{\delta N_{x}}\left(a_{x}p\left[N_{x},s\right]\right)+\frac{1}{2}\int\text{d}^{4}y\,n_{y}\,\frac{\delta}{\delta N_{x}}\frac{\delta}{\delta N_{y}}\left(b_{x}b_{y}p\left[N_{x},s\right]\right)\,k_{\varepsilon}\left(x,y\right)\right]\\&=-\int\text{d}^{4}x\,\text{d}N_{x}\,m_{x}\,f\frac{\delta}{\delta N_{x}}\left[\left(a_{x}p\left[N_{x},s\right]\right)-\frac{1}{2}\int\text{d}^{4}y\,n_{y}\,\frac{\delta}{\delta N_{y}}\left(b_{x}b_{y}p\left[N_{x},s\right]\right)\,k_{\varepsilon}\left(x,y\right)\right]\\&\\&=-\int\text{d}^{4}x\,\text{d}N_{x}\,m_{x}\,f\frac{\delta}{\delta N_{x}}\left[\left(a_{x}p\left[N_{x},s\right]\right)-\frac{1}{2}\int\text{d}^{4}y\text{d}^{4}z\,n_{y}\,\frac{\delta}{\delta N_{y}}\left(b_{x}b_{z}p\left[N_{x},s\right]\right)\,k_{\varepsilon}\left(x,y\right)k_{\varepsilon}\left(y,z\right)\right]\,.
    \end{split}
\end{equation}
We may write at this point the Fokker-Planck equation at equilibrium (i.e. for stationary $p[N]$), using the property $k_{\varepsilon}\left(x,x\right)=\alpha^2$, namely

\begin{equation}
    \begin{split}
&a_{x}p\left[N_{x}\right]-\frac{e^{\imath\gamma}\alpha}{4}\left(2n_{x}\,N_{x}p\left[N_{x}\right]+n_{x}N_{x}^{2}\frac{\delta}{\delta N_{x}}p\left[N_{x}\right]\right)=0\,,\\
    \end{split}
\end{equation}        
which through a series of manipulations leads to        
 \begin{equation}
    \begin{split}       
        &\frac{\delta}{\delta N_{x}}\log\left(p\left[N_{x}\right]\right)=\frac{4}{e^{\imath\gamma}\alpha n_{x}N_{x}^{2}}\left[a_{x}-\frac{e^{\imath\gamma}\alpha}{2}n_{x}\,N_{x}\right]\,.\\
\end{split}
\end{equation}   
This letter can be finally integrated, considering arbitrary complex constants of integration $A$ and $B$, according to 
\begin{equation}
    \begin{split}
&\int\prod_{x}\delta\log\left(p\left[N_{x}\right]\right)=\int\prod_{x}\text{d}N_{x}\frac{4}{e^{\imath\gamma}\alpha n_{x}N_{x}^{2}}\left[a_{x}-\frac{e^{\imath\gamma}\alpha}{2}n_{x}\,N_{x}\right]\,,\\
            \end{split}
\end{equation}
yielding the solution 
\begin{equation}
    \begin{split}
        &\log\left(p\left[N_{x}\right]\right)-A=\int^{N_{x}}\prod_{y}\text{d}N_{y}\frac{4}{e^{\imath\gamma}\alpha n_{y}N_{y}^{2}}\left[a_{y}-\frac{e^{\imath\gamma}\alpha}{2}n_{y}\,N_{y}\right]\,,\\
         \end{split}
\end{equation}
or equivalently,    
\begin{equation} \label{sand}
    \begin{split}  
&p\left[N_{x}\right]=B\exp\left\{ \int^{N_{x}}\prod_{y}\text{d}N_{y}\frac{4}{e^{\imath\gamma}\alpha N_{y}^{2}}\left[a_{y}-\frac{e^{\imath\gamma}\alpha}{2}\,N_{y}\right]\right\} =B\exp\left[\frac{\imath}{\hbar}S_{0}\right] \,,
    \end{split}
\end{equation}
where the last equality follows from a general result that holds for the Euclidean path-integral measure, at the equilibrium of the Fokker-Planck equation, namely that the solution is proportional to exponential of the action.

\section{WKB approximation}
\noindent 
Having derived a solution for the stationary limit of the Fokker-Planck equation, in stead taking the steady-state limit, we now substitute the semi-classical solution for the metric into Eq.~\eqref{sand}. This will automatically reshuffle the action along the equilibrium limit. For this purpose, we resort to the WKB approximation, which for the wave-functional reads
\begin{equation}
\Psi=\exp\left[\frac{\imath}{\hbar}S_{0}\right]=\exp\left[\int^{N_{x}}\prod_{y}\text{d}N_{y}\frac{4}{e^{\imath\gamma}\alpha N_{y}^{2}}\left[a_{y}-\frac{e^{\imath\gamma}\alpha}{2}\,N_{y}\right]\right]\,,
\end{equation}
i.e. it is already found within the same form than in Eq.~\eqref{sand}. This is not surprising, since we already commented that this behaviour is ensured from general arguments that concern the Euclidean path integral measure at equilibrium. \\

The stochastic equation for the wave-functional reads
\begin{equation}
    \begin{split} & \text{d}\Psi=\int\text{d}^{4}x\,m_{x}\,\frac{\delta\Psi}{\delta N_{x}}a_{x}\text{d}s+\int\text{d}^{4}x\,m_{x}\,\frac{\delta\Psi}{\delta N_{x}}b_{x}\text{d}W_{x}
 +\frac{1}{2}\int\text{d}^{4}x\,m_{x}\,\int\text{d}^{4}y\,n_{y}\,b_{x}b_{y}\,k_{\varepsilon}\left(x,y\right)\frac{\delta}{\delta N_{x}}\frac{\delta}{\delta N_{y}}\Psi\text{d}s\,.
\end{split}
\end{equation}

Using the WKB form of $\Psi$, we recover, applying the functional derivatives in $N$, the relation 
 \begin{equation}
    \begin{split}       
 &\text{d}\Psi=\left[\int\text{d}^{4}x\,m_{x}\,\frac{\imath}{\hbar}\frac{\delta S_{0}}{\delta N_{x}}a_{x}\text{d}s+\int\text{d}^{4}x\,m_{x}\,\frac{\imath}{\hbar}\frac{\delta S_{0}}{\delta N_{x}}b_{x}\text{d}W_{x}\right]\Psi\\
       &+\left[\frac{1}{2}\int\text{d}^{4}x\,m_{x}\,\int\text{d}^{4}y\,n_{y}\,b_{x}b_{y}\,k_{\varepsilon}\left(x,y\right)\left(-\frac{1}{\hbar^{2}}\frac{\delta S_{0}}{\delta N_{y}}\frac{\delta S_{0}}{\delta N_{x}}+\frac{\imath}{\hbar}\frac{\delta^{2}S_{0}}{\delta N_{y}\delta N_{x}}\right)\text{d}s\right]\Psi\,,\\
\end{split}
\end{equation}  
in which we finally substitute the expression we determined in Eq.~\eqref{sand} for $S_0$, 
and finally derive
\begin{equation}\label{mare}
    \begin{split}
&\text{d}\Psi=\left[\int\text{d}^{4}x\,m_{x}\,\frac{4}{e^{\imath\gamma}\alpha N_{x}^{2}}\left[a_{x}-\frac{e^{\imath\gamma}\alpha}{2}\,N_{x}\right]a_{x}\text{d}s+\int\text{d}^{4}x\,m_{x}\,\frac{4}{e^{\imath\gamma}\alpha N_{x}^{2}}\left[a_{x}-\frac{e^{\imath\gamma}\alpha}{2}\,N_{x}\right]b_{x}\text{d}W_{x}\right]\Psi\\&+\left[\frac{1}{2}\int\text{d}^{4}x\,m_{x}\,\int\text{d}^{4}y\,n_{y}\,b_{x}b_{y}\,k_{\varepsilon}\left(x,y\right)\left(\frac{16}{e^{2\imath\gamma}\alpha^{2}}\frac{1}{N_{x}^{2}N_{y}^{2}}\left[a_{y}-\frac{e^{\imath\gamma}\alpha}{2}\,N_{y}\right]\left[a_{x}-\frac{e^{\imath\gamma}\alpha}{2}\,N_{x}\right]\right)\text{d}s\right]\Psi\\&+\left[\frac{1}{2}\int\text{d}^{4}x\,m_{x}\,\int\text{d}^{4}y\,n_{y}\,b_{x}b_{y}\,k_{\varepsilon}\left(x,y\right)\left(\left(-\frac{8\delta^{4}\left(x-y\right)}{e^{\imath\gamma}\alpha N_{x}^{3}}\left[a_{x}-\frac{e^{\imath\gamma}\alpha}{2}\,N_{x}\right]+\frac{4}{e^{\imath\gamma}\alpha N_{x}^{2}}\left[\frac{\delta a_{x}}{\delta N_{y}}-\frac{e^{\imath\gamma}\alpha}{2}\delta^{4}\left(x-y\right)\right]\right)\right)\text{d}s\right]\Psi\,.
    \end{split}
\end{equation}

\section{Non-relativistic limit}\label{NRL}
\noindent 
We are now ready to consider the non-relativistic limit, in the WKB approximated solution. We first remind the expression for the non-trivial components of the metric that 
\begin{equation}
    g_{00}=-1-\frac{2\phi}{c^{2}},\qquad \qquad N=\sqrt{1+\frac{2\phi}{c^{2}}}\simeq1+\frac{\phi}{c^{2}}\,.
\end{equation}
Recalling the definition of $a$, slightly reshuffled, we write
\begin{equation}
    \begin{split} & a_{x}=-\frac{\imath}{2}\left[2\left(N_{x}\left(\nabla\log N_{x}\right)^{2}+\nabla^{2}N_{x}\right)+\kappa N_{x}\rho_{x}\left(3\sigma T+1\right)\right]-\frac{\alpha}{4}e^{\imath\gamma}N_{x}\,,\\
\frac{\delta a_{x}}{\delta N_{y}} & =-\frac{\imath}{2}\left[2\left(\left(\nabla\log N_{x}\right)^{2}\delta^{4}\left(x-y\right)+2N_{x}\nabla\log N_{x}\nabla\frac{\delta^{4}\left(x-y\right)}{N_{x}}+\nabla^{2}\delta^{4}\left(x-y\right)\right)\right]\\
 &\ \  \! \ \ -\frac{\imath}{2}\left[\kappa\rho_{x}\left(3\sigma T+1\right)+\frac{\alpha}{2}e^{\imath\gamma-\pi/2}\right]\,.
\end{split}
\end{equation}
We then approximate the terms $\nabla^2 N\simeq \nabla^2 \phi=\kappa \rho /2$ , hence using the classical limit, and consider the approximations $N\simeq 1$ and $N^{-2}\simeq 1$. The first term in the first line of Eq.~\eqref{mare} can be reshuffled according to a series of passages.
\begin{equation}
    \begin{split} 
    & \Psi \int\text{d}^{4}x\,m_{x}\,\frac{4}{e^{\imath\gamma}\alpha N_{x}^{2}}\left[a_{x}-\frac{e^{\imath\gamma}\alpha}{2}\,N_{x}\right]a_{x}\text{d}s \\
= &\Psi\int\text{d}^{4}x\,m_{x}\,\frac{4}{e^{\imath\gamma}\alpha N_{x}^{2}}\left[-\imath\nabla_{x}^{2}\phi_{x}-\frac{\imath}{2}N_{x}\kappa\rho_{x}\left(3\sigma T+1\right)-\frac{3}{4}\alpha e^{\imath\gamma}N_{x}\right]
 \left[-\imath\nabla^{2}\phi_{x}-\frac{\imath}{2}N_{x}\kappa\rho_{x}\left(3\sigma T+1\right)-\frac{\alpha}{4}N_{x}e^{\imath\gamma}\right]\text{d}s\\
= & \Psi\int\text{d}^{4}x\,m_{x}\,\frac{4}{e^{\imath\gamma}\alpha N_{x}^{2}}\left[-\imath\frac{\kappa}{2}\rho_{x}-\frac{\imath}{2}N_{x}\kappa\rho_{x}\left(3\sigma T+1\right)-\frac{3}{4}\alpha e^{\imath\gamma}N_{x}\right]\left[-\imath\frac{\kappa}{2}\rho_{x}-\frac{\imath}{2}N_{x}\kappa\rho_{x}\left(3\sigma T+1\right)-\frac{\alpha}{4}N_{x}e^{\imath\gamma}\right]\text{d}s\\
= & \Psi\int\text{d}^{4}x\,m_{x}\,\frac{4}{e^{\imath\gamma}\alpha}\left[-\frac{\imath}{2}\kappa\rho_{x}\left(3\sigma T+2\right)-\frac{3}{4}\alpha e^{\imath\gamma}\right]
 \left[-\frac{\imath}{2}\kappa\rho_{x}\left(3\sigma T+2\right)-\frac{\alpha}{4}e^{\imath\gamma}\right]\text{d}s\\
= & \Psi\int\text{d}^{4}x\,m_{x}\,\frac{4}{e^{\imath\gamma}\alpha}\left[-\frac{\imath}{2}\kappa\rho_{x}\left(3\sigma T+2\right)\left[-\frac{\imath}{2}\kappa\rho_{x}\left(3\sigma T+2\right)-\frac{\alpha}{4}e^{\imath\gamma}\right]-\frac{3}{4}\alpha e^{\imath\gamma}\left[-\frac{\imath}{2}\kappa\rho_{x}\left(3\sigma T+2\right)-\frac{\alpha}{4}e^{\imath\gamma}\right]\right]\text{d}s\\
= & \Psi\int\text{d}^{4}x\,m_{x}\,\left[-\frac{\imath}{2}\kappa\rho_{x}\left(3\sigma T+2\right)\left[-1\right]-3\left[-\frac{\imath}{2}\kappa\rho_{x}\left(3\sigma T+2\right)-\frac{\alpha}{4}e^{\imath\gamma}\right]\right]\text{d}s\\
= & \Psi\int\text{d}^{4}x\,m_{x}\,\left[\frac{\imath}{2}\kappa\rho_{x}\left(3\sigma T+2\right)+3\frac{\imath}{2}\kappa\rho_{x}\left(3\sigma T+2\right)+\frac{\alpha}{4}e^{\imath\gamma}\right]\text{d}s\\
= & \Psi\int\text{d}^{4}x\,m_{x}\,\left[2\imath\kappa\rho_{x}\left(2+3\sigma T\right)+\frac{\alpha}{4}e^{\imath\gamma}\right]\text{d}s\,.
\end{split}
\end{equation}
The second term in the first line of Eq.~\eqref{mare} reads
\begin{equation}
    \begin{split} & \Psi\int\text{d}^{4}x\,m_{x}\,\frac{4}{e^{\imath\gamma}\alpha N_{x}^{2}}\left[a_{x}-\frac{e^{\imath\gamma}\alpha}{2}\,N_{x}\right]b_{x}\text{d}W_{x}\\
= & \Psi\int\text{d}^{4}x\,m_{x}\,\frac{4}{e^{\imath\gamma}\alpha N_{x}^{2}}\left[-\frac{\imath}{2}\left[2\left(N_{x}\left(\nabla\log N_{x}\right)^{2}+\nabla^{2}N_{x}\right)+\kappa N_{x}\rho_{x}\left(3\sigma T+1\right)\right]-\frac{\alpha}{4}e^{\imath\gamma}N_{x}-\frac{e^{\imath\gamma}\alpha}{2}\,N_{x}\right]e^{\imath\gamma/2}\sqrt{\frac{\alpha}{2}}N_{x}\text{d}W_{x}\\
= & \Psi e^{\imath\gamma/2}\sqrt{\frac{\alpha}{2}}\int\text{d}^{4}x\,m_{x}\,\frac{4}{e^{\imath\gamma}\alpha N_{x}^{2}}\left[-\imath\left(N_{x}^{2}\left(\nabla\log N_{x}\right)^{2}+N_{x}\nabla^{2}N_{x}\right)-\frac{\imath}{2}\kappa N_{x}^{2}\rho_{x}\left(3\sigma T+1\right)-\frac{\alpha}{4}e^{\imath\gamma}N_{x}^{2}-\frac{e^{\imath\gamma}\alpha}{2}\,N_{x}^{2}\right]\text{d}W_{x}\\
= & \Psi e^{\imath\gamma/2}\sqrt{\frac{1}{2\alpha}}\int\text{d}^{4}x\,m_{x}\,4\left[-\imath\frac{e^{-\imath\gamma}}{N_{x}}\frac{\kappa}{2}\rho_{x}-\frac{\imath e^{-\imath\gamma}}{2}\kappa\rho_{x}\left(3\sigma T+1\right)-\frac{3}{4}\alpha\right]\text{d}W_{x}\\
= & \Psi e^{-\imath\gamma/2}\sqrt{\frac{1}{2\alpha}}\int\text{d}^{4}x\,m_{x}\,\left[-2\imath\kappa\rho_{x}\left(3\sigma T+2\right)-3e^{\imath\gamma}\alpha\right]\text{d}W_{x}\\
= & \Psi e^{-\imath\gamma/2}\sqrt{\kappa}\sqrt{\frac{\kappa}{2\alpha\kappa}}\int\text{d}^{4}x\,m_{x}\,\left[-2\imath\sqrt{\kappa}\rho_{x}\left(3\sigma T+2\right)-3e^{\imath\gamma}\frac{\alpha}{\sqrt{\kappa}}\right]\text{d}W_{x}\\
= & \Psi e^{-\imath\gamma/2}\sqrt{\kappa}\sqrt{\frac{1}{2\alpha\kappa}}\int\text{d}^{4}x\,m_{x}\,\left[-2\imath\kappa\rho_{x}\left(3\sigma T+2\right)-3e^{\imath\gamma}\alpha\right]\text{d}W_{x}\\
= & \Psi e^{-\imath\gamma/2}\sqrt{\kappa}\sqrt{\frac{\kappa}{2\alpha}}\int\text{d}^{4}x\,m_{x}\,\left[-2\imath\rho_{x}\left(3\sigma T+2\right)-3e^{\imath\gamma}\frac{\alpha}{\kappa}\right]\text{d}W_{x}\,.
\end{split}
\end{equation}
The third term, in the second line of Eq.~\eqref{mare}, reads
\begin{equation}
    \begin{split}
        &+\left[\frac{1}{2}\int\text{d}^{4}x\,m_{x}\,\int\text{d}^{4}y\,n_{y}\,b_{x}b_{y}\,k_{\varepsilon}\left(x,y\right)\left(\frac{16}{e^{2\imath\gamma}\alpha^{2}}\frac{1}{N_{x}^{2}N_{y}^{2}}\left[a_{y}-\frac{e^{\imath\gamma}\alpha}{2}\,N_{y}\right]\left[a_{x}-\frac{e^{\imath\gamma}\alpha}{2}\,N_{x}\right]\right)\text{d}s\right]\Psi\\&+\left[\frac{1}{2}\int\text{d}^{4}x\,m_{x}\,n_{x}\,b_{x}^{2}\,\left(-\frac{8}{e^{\imath\gamma}\alpha N_{x}^{3}}\left[a_{x}-\frac{e^{\imath\gamma}\alpha}{2}\,N_{x}\right]\right)\text{d}s\right]\Psi\\&+\left[\frac{1}{2}\int\text{d}^{4}x\,m_{x}\,n_{x}\,b_{x}^{2}\left(\left(\frac{4}{e^{\imath\gamma}\alpha N_{x}^{2}}\left[-\frac{1}{2}\left[\imath\kappa\rho_{x}\left(3\sigma T+1\right)-\frac{\alpha}{2}e^{\imath\gamma}\right]\right]\right)\right)\text{d}s\right]\Psi\\&+\left[\frac{1}{2}\int\text{d}^{4}x\,m_{x}\,n_{x}\,b_{x}^{2}\left(\left(\frac{4}{e^{\imath\gamma}\alpha N_{x}^{2}}\left[-\frac{e^{\imath\gamma}\alpha}{2}\right]\right)\right)\text{d}s\right]\Psi\\
     =&+\left[\frac{1}{2}\int\text{d}^{4}x\,m_{x}\,\int\text{d}^{4}y\,n_{y}\,b_{x}b_{y}\,k_{\varepsilon}\left(x,y\right)\left(\frac{16}{e^{2\imath\gamma}\alpha^{2}}\frac{1}{N_{x}^{2}N_{y}^{2}}\left[a_{y}-\frac{e^{\imath\gamma}\alpha}{2}\,N_{y}\right]\left[a_{x}-\frac{e^{\imath\gamma}\alpha}{2}\,N_{x}\right]\right)\text{d}s\right]\Psi\\&+\left[\frac{1}{2}\int\text{d}^{4}x\,m_{x}\,n_{x}\,\left(3\imath\kappa\rho_{x}\left(\sigma T+1\right)+\frac{5}{2}e^{\imath\gamma}\alpha\right)\text{d}s\right]\Psi\,,
    \end{split}
\end{equation}
where the first line reads
\begin{equation}
    \begin{split} & +\frac{1}{2}\int\text{d}^{4}x\,m_{x}\,\int\text{d}^{4}y\,n_{y}\,b_{x}b_{y}\,k_{\varepsilon}\left(x,y\right)\frac{16}{e^{2\imath\gamma}\alpha^{2}}\text{d}s\Psi\\
 & \times\frac{1}{N_{x}^{2}N_{y}^{2}}\left[\left[-\frac{\imath}{2}\left[2\left(N_{y}\left(\nabla\log N_{y}\right)^{2}+\nabla_{y}^{2}N_{y}\right)+\kappa N_{y}\rho_{y}\left(3\sigma T+1\right)\right]-\frac{\alpha}{4}e^{\imath\gamma}N_{y}\right]-\frac{e^{\imath\gamma}\alpha}{2}\,N_{y}\right]\\
 & \times\left[\left[-\frac{\imath}{2}\left[2\left(N_{x}\left(\nabla\log N_{x}\right)^{2}+\nabla_{x}^{2}N_{x}\right)+\kappa N_{x}\rho_{x}\left(3\sigma T+1\right)\right]-\frac{\alpha}{4}e^{\imath\gamma}N_{x}\right]-\frac{e^{\imath\gamma}\alpha}{2}\,N_{x}\right]\\
\\
= & +\frac{1}{2}\int\text{d}^{4}x\,m_{x}\,\int\text{d}^{4}y\,n_{y}\,e^{\imath\gamma}\frac{\alpha}{2}N_{x}N_{y}\,k_{\varepsilon}\left(x,y\right)\frac{16}{e^{2\imath\gamma}\alpha^{2}}\text{d}s\Psi\\
 & \times\frac{1}{N_{y}}\left[-\frac{\imath}{2}\left[\frac{1}{N_{y}}\kappa\rho_{y}+\kappa\rho_{y}\left(3\sigma T+1\right)\right]-\frac{\alpha}{4}e^{\imath\gamma}-\frac{e^{\imath\gamma}\alpha}{2}\right] \frac{1}{N_{x}}\left[-\frac{\imath}{2}\left[\frac{1}{N_{x}}\kappa\rho_{x}+\kappa\rho_{x}\left(3\sigma T+1\right)\right]-\frac{\alpha}{4}e^{\imath\gamma}-\frac{e^{\imath\gamma}\alpha}{2}\right]\\
\\
= & +\frac{1}{2}\int\text{d}^{4}x\,m_{x}\,\int\text{d}^{4}y\,n_{y}\,\frac{1}{2}\,k_{\varepsilon}\left(x,y\right)\frac{16}{e^{\imath\gamma}\alpha}\text{d}s\Psi\\
 & \times\left[-\frac{\imath}{2}\left[\frac{1}{N_{y}}\kappa\rho_{y}+\kappa\rho_{y}\left(3\sigma T+1\right)\right]-\frac{3}{4}\alpha e^{\imath\gamma}\right] \left[-\frac{\imath}{2}\left[\frac{1}{N_{x}}\kappa\rho_{x}+\kappa\rho_{x}\left(3\sigma T+1\right)\right]-\frac{3}{4}\alpha e^{\imath\gamma}\right]\,.
\end{split}
\end{equation}
Finally, the last line of Eq.~\eqref{mare} can be dropped, due to the presence of the Dirac delta function.\\

Combining the three non-vanishing terms together, after a few manipulations we obtain 
\begin{eqnarray} 
\text{d} \Psi \!&=&\! \left\{\int\text{d}^{4}x\,m_{x}\,\left[2\imath\kappa\rho_{x}\left(2+3\sigma T\right)+\frac{\alpha}{4}e^{\imath\gamma}\right]\text{d}s \right\}\, \Psi - \left\{    \frac{e^{-\imath\frac{\gamma}{2}}}{\sqrt{2 \alpha}} \!\! \int \!\! \text{d}^{4}x\,m_{x}\! \left[2\imath\kappa \rho_{x}\left(3\sigma T+2\right) \!+\! 3\alpha e^{\imath\gamma}\right] \! \text{d}W_{x} \! \right\} \! \Psi \nonumber\\
&+& \Bigg\{\frac{1}{2}\int\text{d}^{4}x\,m_{x}\,\int\text{d}^{4}y\,n_{y}\,\frac{1}{2}\,k_{\varepsilon}\left(x,y\right)\frac{1}{e^{\imath\gamma}\alpha}\text{d}s  \left[2 \imath \kappa\rho_{x}\left(2\!+\! 3\sigma T\right)\!+\! 3 \alpha e^{\imath\gamma}\right] 
  \left[2 \imath \kappa\rho_{y}\left(2\!+\! 3\sigma T\right)\!+\!3 \alpha e^{\imath\gamma}\right] 
 \! \Bigg\}\Psi\,. \nonumber
\end{eqnarray}
The low temperature limit is accomplished by dropping all terms in $\sigma T$ out of these equations, which leads to
\begin{eqnarray} 
\text{d} \Psi \!&=&\! \left\{\int\text{d}^{4}x\,m_{x}\,\left[4\imath\kappa\rho_{x} +\frac{\alpha}{4}e^{\imath\gamma}\right]\text{d}s \right\}\, \Psi - \left\{    \frac{e^{-\imath\frac{\gamma}{2}}}{\sqrt{2 \alpha}} \!\! \int \!\! \text{d}^{4}x\,m_{x}\! \left[4\imath\kappa \rho_{x} \!+\! 3\alpha e^{\imath\gamma}\right] \! \text{d}W_{x} \! \right\} \! \Psi \nonumber\\
&+& \Bigg\{\frac{1}{2}\int\text{d}^{4}x\,m_{x}\,\int\text{d}^{4}y\,n_{y}\,\frac{1}{2}\,k_{\varepsilon}\left(x,y\right)\frac{1}{e^{\imath\gamma}\alpha}\text{d}s  \left[4 \imath \kappa\rho_{x} \!+\! 3 \alpha e^{\imath\gamma}\right] 
  \left[4 \imath \kappa\rho_{y} \!+\!3 \alpha e^{\imath\gamma}\right] 
 \! \Bigg\}\Psi. \nonumber
\end{eqnarray}

A small phase expansion can be also considered. A privileged phase is $\gamma=\pi/2$, which corresponds to a saddle point of the stationary phase of the Fokker-Planck equation. Expanding for small $\gamma-\pi/2$, considering only the lowest (zeroth) order of this expansion, further simplifies the equation into

\begin{eqnarray} 
\text{d} \Psi \!&=&\! \left\{ \imath \int\text{d}^{4}x\,m_{x}\,\left[4 \kappa\rho_{x} +\frac{\alpha}{4} \right]\text{d}s \right\}\, \Psi - \left\{   \imath  e^{-\imath\frac{\pi}{4}} \!\! \int \!\! \text{d}^{4}x\,m_{x}\! \left[4 \kappa \rho_{x} \!+\! 3\alpha \right] \! \text{d}\widetilde{W}_{x} \! \right\} \! \Psi \nonumber\\
&-& \Bigg\{\frac{\alpha}{2}\int\text{d}^{4}x\,m_{x}\,\int\text{d}^{4}y\,n_{y}\,\widetilde{k}_{\varepsilon}\left(x,y\right) \text{d}s  \left[4 \kappa\rho_{x} \!+\! 3 \alpha \right] 
  \left[4  \kappa\rho_{y} \!+\!3 \alpha \right] 
 \! \Bigg\}\Psi, \nonumber
\end{eqnarray}
in which the Wiener process $\text{d}\widetilde{W}_x = \text{d}{W}_x/\sqrt{2 \alpha}$, with dimension of energy square, and the dimensionless kernel $\widetilde{k}_{\varepsilon}\left(x,y\right)={k}_{\varepsilon}\left(x,y\right)/(2\alpha^2)$ have been introduced. 

This latter relation can be recast, introducing the definitions of Hamiltonian density function $\mathcal{H}$ and jump density function $\mathcal{L}$,
\begin{equation}
\mathcal{H}= 4 \kappa\rho +\frac{\alpha}{4} \,, \qquad \mathcal{L} = \mathcal{H} + \frac{15}{4} \alpha \,,
\end{equation}
into 
\begin{eqnarray} \label{anal}
\text{d} \Psi \!&=&\! \left\{ \imath \int\text{d}^{4}x\,m_{x}\,\mathcal{H}_x \text{d}s \right\}\, \Psi - \left\{   \imath  e^{-\imath\frac{\pi}{4}} \!\! \int \!\! \text{d}^{4}x\,m_{x} \mathcal{L}_x  \text{d}\widetilde{W}_{x} \! \right\} \! \Psi - \Bigg\{\frac{\alpha}{2}\int\text{d}^{4}x\,m_{x}\,\int\text{d}^{4}y\,n_{y}\,\widetilde{k}_{\varepsilon}\left(x,y\right) \mathcal{L}_x \mathcal{L}_y \text{d}s \! \Bigg\}\Psi\,. \label{sabba}
\end{eqnarray}

\section{Gorini-Kossakowski-Lindblad-Sudarshan equation} \label{SecGKLS}
\noindent 
Eq.~\eqref{sabba} enables to derive the Gorini-Kossakowski-Lindblad-Sudarshan equation directly from the context of the SRF. In order to derive this result, we first notice, as already specified by Parisi and Wu in \cite{Parisi:1980ys}, that the kernel is connected to the expectation value of the product of two copies of the stochastic noise at different values of the thermal time and in two different space-time points, i.e.
\begin{equation}
\langle \tilde{\eta}(x, s) \tilde{\eta}(x', s') \rangle_{\tilde{\eta}} = \widetilde{k}_{\varepsilon}(x,x') \, \delta(s,s')\,.
\end{equation}
Consistently, for the Wiener process $\text{d}\widetilde{W}_x$ it holds 
\begin{equation}
\text{d}\widetilde{W}_x= \tilde{\eta}_x \, ds\,\alpha^{\frac{3}{2}}\,.
\end{equation}
Thus both the second and the third term of Eq.~\eqref{anal} vanish in the $\alpha \rightarrow 0$ limit, i.e. in the zero cosmological constant limit --- this is also the zero stochastic gravitational noise limit, as discussed in \cite{Lulli:2021bme} --- because of their analytical dependence in $\alpha$. This behaviour ensures that the relaxation of the quantum system, due to the stochastic noise of gravity and interpreted as collapse of the wave-function of the quantum system, disappears in absence of noise, leaving the system to evolve coherently, according to the Schr\"odinger equation.\\

We can start considering how to reconstruct the dynamics of the density matrix from the SRF equation of the wave-functional $\Psi$. For this purpose, we first recall what has been already clarified in \cite{Lulli:2021bme}, namely that the thermal time is monotonic in the proper time, and thus it can be used to describe the dynamics of a system. Then, after having considered in the previous steps both the non-relativistic limit and the semi-classical WKB approximation, it is necessary now to go back to a first quantization of the system, which has been recognized to undergo the classical dynamics dictated by Eq.~\eqref{sabba}. This passage implies to turn the density functions $\mathcal{H}$ and $\mathcal{L}$ into two density operators acting on the Hilbert space of states that describe the quantum system under scrutiny. We then represent the wave-functional $\Psi$, previously considered in the semi-relativistic limit, as a wave-function, or equivalently a ket $| \Psi\rangle$. Consequently, we promote Eq.~\eqref{sabba} at the quantum level, as 
\begin{eqnarray} 
|\text{d}  \Psi\rangle \!&=&\! \left\{ \imath \int\text{d}^{4}x\,m_{x}\,\mathcal{H}_x \text{d}s \right\}\, |\Psi \rangle - \left\{   \imath  e^{-\imath\frac{\pi}{4}} \!\! \int \!\! \text{d}^{4}x\,m_{x} \mathcal{L}_x  \text{d}\widetilde{W}_{x} \! \right\} \! |\Psi\rangle - \Bigg\{\frac{1}{2}\int\text{d}^{4}x\,m_{x}\,\int\text{d}^{4}y\,n_{y}\,\widetilde{k}_{\varepsilon}\left(x,y\right) \mathcal{L}_x \mathcal{L}_y \text{d}s \! \Bigg\} |\Psi \rangle,  \ \ \ \ \label{sabba2}
\end{eqnarray}
from which it follows
\begin{eqnarray} 
 \langle \text{d} \Psi |  \!&=&\!  - \langle \Psi | \left\{  \imath \int\text{d}^{4}x\,m_{x}\,\mathcal{H}_x \text{d}s \right\} + \langle \Psi | \left\{   \imath  e^{\imath\frac{\pi}{4}} \!\! \int \!\! \text{d}^{4}x\,m_{x} \mathcal{L}_x  \text{d}\widetilde{W}_{x} \! \right\}  - \langle \Psi | \Bigg\{\frac{1}{2}\int\text{d}^{4}x\,m_{x}\,\int\text{d}^{4}y\,n_{y}\,\widetilde{k}_{\varepsilon}\left(x,y\right) \mathcal{L}_y \mathcal{L}_x \text{d}s \! \Bigg\}, \ \ \ \ \label{sabba3}
\end{eqnarray}
having used the fact that both the Hamiltonian density operator $\mathcal{H}$ and jump density operator $\mathcal{L}$ are hermitian, namely that 
\begin{equation}
\mathcal{H}= \mathcal{H}^\dagger\,, \qquad \qquad \mathcal{L}= \mathcal{L}^\dagger\,.
\end{equation}
Now, being $|\Psi\rangle$ a generic state, we consider the density matrix $\rho_\Psi \equiv |\Psi\rangle  \langle \Psi |$ constructed out of it, and hence the variation at $\mathcal{O} (ds)$, namely the first order in $\text{d}s$, of its value averaged over the stochastic noise distribution, i.e. 
\begin{eqnarray}
\left\langle \text{d} \rho_\Psi \Big|_{\mathcal{O} (ds)}  \right\rangle_{\tilde{\eta}} &=& \left\langle \big[ \left( |\Psi\rangle + | \text{d} \Psi\rangle  \right) \left( \langle \Psi | +  \langle \text{d} \Psi | \right)\big] \Big|_{\mathcal{O} (ds)} \right\rangle_{\tilde{\eta}} \nonumber\\
&=&  \Big\langle   |\Psi\rangle  \langle \text{d} \Psi |    \Big\rangle_{\tilde{\eta}} \big|_{\mathcal{O} (ds)} + \Big\langle   |\text{d} \Psi\rangle  \langle  \Psi |    \Big\rangle_{\tilde{\eta}} \big|_{\mathcal{O} (ds)} + \Big\langle   |\text{d} \Psi\rangle  \langle \text{d}  \Psi |    \Big\rangle_{\tilde{\eta}} \big|_{\mathcal{O} (ds)}\,, \label{sabbade}
\end{eqnarray}
where in the last equality of \eqref{sabbade} an additional term to the standard Leibnitz rule is required, due to the co-presence of both order $\mathcal{O}(\text{d} s^{\frac{1}{2}})$ and order $\mathcal{O}(\text{d} s)$ terms in the expression for the SDE of $|\Psi\rangle$. On the other hand, the co-presence of these latter terms is implied by the choice of the It\^o calculus. 

Considering that 
\begin{eqnarray}
\left\langle \text{d} \widetilde{W} \right\rangle_{\tilde{\eta}} =0\,, \qquad \qquad \left\langle \text{d} \widetilde{W}_x \text{d} \widetilde{W}_y  \right\rangle_{\tilde{\eta}} = \widetilde{k}_{\varepsilon}\left(x,y\right) \text{d} s\,, 
\end{eqnarray}
we arrive to the final expression
\begin{eqnarray} \label{GKLSgen}
\frac{\text{d}}{\text{d} s}\rho_\Psi \equiv  \left\langle \frac{\text{d}}{\text{d} s}\rho_\Psi \right\rangle_{\tilde{\eta}} =  \imath \int d^4x \, m_x \left[ \mathcal{H}_x, \rho_\Psi \right] &+& \int d^4x\,  m_x  \int d^4y\,  n_y \, \widetilde{k}_{\varepsilon}\left(x,y\right) \, \mathcal{L}_x \rho_\Psi \mathcal{L}_y^\dagger \nonumber\\
&-&\frac{1}{2}  \int d^4x\,  m_x  \int d^4y\,  n_y \, \widetilde{k}_{\varepsilon}\left(x,y\right)\,  \{\mathcal{L}_x^\dagger \mathcal{L}_y, \rho_\Psi \}
\,.
\end{eqnarray}
This expression is more general than the GKLS form, as it does not imply a specific choice of the kernel, hence a specific choice of the stochastic noise. 

When the kernel, in the $\varepsilon \to 0 $ limit, converges to a white noise distribution, then Eq.~\eqref{GKLSgen} provides exactly the GKLS form, namely
\begin{eqnarray} \label{GKLS}
\frac{\text{d}}{\text{d} s}\rho_\Psi \equiv  \left\langle \frac{\text{d}}{\text{d} s}\rho_\Psi \right\rangle_{\tilde{\eta}} = \imath \left[ H, \rho_\Psi \right] + L\rho_\Psi L^\dagger -\frac{1}{2} \{L^\dagger L, \rho_\Psi \}
\,,
\end{eqnarray}
provided that the Hamiltonian operator $H=\int d^3x \mathcal{H}$ and the jump operator $L=\int d^3x \mathcal{L}$ have been introduced and that, by a proper choice of the smearing functions $n_x$ and $m_x$ that restricts the time coordinate to the proper time, in turn related to the thermal time $s$ through the change of coordinates specified in \cite{Lulli:2021bme}, one can write
\begin{eqnarray} \label{GKLS}
\int\text{d}^{4}x\,m_{x}\,\mathcal{H}_x \text{d}s =  {H}_s \text{d}s\,, \qquad \qquad \int\text{d}^{4}x\,m_{x}\,\mathcal{L}_x \text{d}s = \,{L}_s \text{d}s\,.
\end{eqnarray}
\bibliographystyle{apsrev4-2}
\bibliography{arXiv_version}

\end{document}